\documentclass{elsart}
\usepackage{epsfig,amssymb}

\begin{document}
\newcommand{\beqn}{\begin{equation}}
\newcommand{\eeqn}{\end{equation}}
\newcommand{\etag}{\eta\gamma}
\newcommand{\ee}{e^+e^-}
\newcommand{\pipi}{\pi^+\pi^-}
\newcommand{\ompi}{\omega\pi^0}
\newcommand{\pig}{\pi^0\gamma}
\newcommand{\pipig}{\pi^0\pig}
\newcommand{\trpi}{\pi^+\pi^-\pi^0}
\newcommand{\rhop}{\rho\prime}
\newcommand{\rhopp}{\rho\prime\prime}
\newcommand{\etapig}{\eta\pi^0\gamma}

\begin{frontmatter}
\date{}

\title{\large \bf \boldmath 
Study of the Process $\ee\to\ompi\to\pipig$ in c.m. Energy Range
920--1380~MeV at CMD-2}

\author[BINP]{R.R.~Akhmetshin},
\author[BINP,NGU]{V.M.~Aulchenko},
\author[BINP]{V.Sh.~Banzarov},
\author[PITT]{A.~Baratt},
\author[BINP,NGU]{L.M.~Barkov},
\author[BINP]{S.E.~Baru},
\author[BINP]{N.S.~Bashtovoy},
\author[BINP,NGU]{A.E.~Bondar},
\author[BINP]{D.V.~Bondarev},
\author[BINP]{A.V.~Bragin},
\author[BINP,NGU]{S.I.~Eidelman},
\author[BINP]{D.A.~Epifanov},
\author[BINP,NGU]{G.V.~Fedotovitch},
\author[BINP]{N.I.~Gabyshev},
\author[BINP]{D.A.~Gorbachev},
\author[BINP]{A.A.~Grebeniuk}, 
\author[BINP]{D.N.~Grigoriev},
\author[YALE]{V.W.~Hughes}~\footnote{deceased},
\author[BINP]{F.V.~Ignatov},
\author[BINP]{S.V.~Karpov},
\author[BINP]{V.F.~Kazanin},
\author[BINP,NGU]{B.I.~Khazin},
\author[BINP,NGU]{I.A.~Koop},
\author[BINP]{P.P.~Krokovny},
\author[BINP,NGU]{A.S.~Kuzmin},
\author[BINP]{I.B.~Logashenko},
\author[BINP]{P.A.~Lukin},
\author[BINP]{A.P.~Lysenko},
\author[BINP]{K.Yu.~Mikhailov},
\author[BINP,NGU]{A.I.~Milstein},
\author[BINP]{I.N.~Nesterenko},
\author[BINP]{V.S.~Okhapkin},
\author[BINP]{A.V.~Otboev},
\author[BINP,NGU]{A.V.~Pak},
\author[BINP,NGU]{E.A.~Perevedentsev},
\author[BINP]{A.A.~Polunin},
\author[BINP]{A.S.~Popov},
\author[BINP]{S.I.~Redin},
\author[BOST]{B.L.~Roberts},
\author[BINP]{N.I.~Root},
\author[BINP]{A.A.~Ruban},
\author[BINP]{N.M.~Ryskulov},
\author[BINP]{A.G.~Shamov}, 
\author[BINP]{Yu.M.~Shatunov},
\author[BINP,NGU]{B.A.~Shwartz},
\author[BINP,NGU]{A.L.~Sibidanov},
\author[BINP]{V.A.~Sidorov}, 
\author[BINP]{A.N.~Skrinsky},
\author[BINP]{I.G.~Snopkov},
\author[BINP,NGU]{E.P.~Solodov},
\author[BINP]{P.Yu.~Stepanov},
\author[PITT]{J.A.~Thompson}, 
\author[BINP]{A.A.~Valishev},
\author[BINP]{Yu.V.~Yudin},
\author[BINP,NGU]{A.S.~Zaitsev},
\author[BINP]{S.G.~Zverev}

\address[BINP]{Budker Institute of Nuclear Physics, 
  Novosibirsk, 630090, Russia}
\address[BOST]{Boston University, Boston, MA 02215, USA}
\address[NGU]{Novosibirsk State University, 
  Novosibirsk, 630090, Russia}
\address[PITT]{University of Pittsburgh, Pittsburgh, PA 15260, USA}
\address[YALE]{Yale University, New Haven, CT 06511, USA}

\begin{abstract}
The cross section of the process $\ee\to\ompi \to \pipig$ has been 
measured in the c.m. energy range 920--1380~MeV with the 
CMD-2 detector. Its energy dependence is well described by the 
interference of the $\rho(770)$ and $\rho(1450)$ mesons decaying to
$\ompi$. Upper limits for the cross sections of the direct processes
$\ee\to \pipig, \etapig$ have been set.
\end{abstract}
\end{frontmatter}
\maketitle

\section{Introduction}
The process $\ee\to\ompi$ is one of the dominant hadronic
processes contributing to the total hadronic cross section at the
c.m. energy between 1 and 2~GeV. The precise measurement of 
its cross section will help to improve the accuracy of the calculation 
of the hadronic contribution to the muon anomalous magnetic 
moment~\cite{g-2} as well as 
check the relations 
between the values of the cross section of the process $\ee\to\ompi$ 
and the differential rate of the 
$\tau^- \to \omega \pi^- \nu_{\tau}$ decay following from
the conservation of the vector current and isospin symmetry~\cite{cvc}.
As one of the important decay modes of the isovector vector states,
it can provide information on the properties of the $\rho$ excitations
as well as clarify the existence  of light exotic
states (hybrids) between 1 and 2~GeV~\cite{hybrid1,hybrid2}.

The dominant decay mode of the $\omega$ meson is that to 
$\pi^+\pi^-\pi^0$ and it is this mode that has been used for the
observation of the process $\ee\to\ompi$ with the complete event 
reconstruction by the DM2~\cite{dm2_ompi}, CMD-2~\cite{cmd2_ompi}
and SND~\cite{snd1} detectors. However, a less probable decay mode
$\omega \to \pig$ is also convenient for the study of the
$\ompi$ final state since it is easier to select than the process
$\ee \to \pi^+\pi^-\pi^0\pi^0$ because of the single contributing
intermediate mechanism. The purely neutral decay mode of this
process has been first studied by ND~\cite{ND}
and recently with much higher statistics by SND~\cite{snd_ompi}. 

In this work we report on the measurement of the cross section 
of the process $\ee\to\ompi\to \pipig$ in the 
c.m. energy range 920--1380~MeV using the CMD-2 detector at the
VEPP-2M $\ee$ collider. 
The preliminary results of this work were published in~\cite{mast}.
We also perform the first search for the direct processes 
$\ee \to \pipig$ and $\ee \to \etapig$ and set upper limits for the
corresponding cross sections.  

\section{Experiment}

The general purpose detector CMD-2 has been described in 
detail elsewhere~\cite{cmddet}. Its tracking system consists of a 
cylindrical drift chamber (DC) and double-layer multiwire proportional 
Z-chamber, both also used for the trigger, and both inside a thin 
(0.38~X$_0$) superconducting solenoid with a field of 1~T. 
The barrel CsI calorimeter with a thickness of 8.1~X$_0$ placed
outside  the solenoid has energy resolution for photons of about
9\% in the energy range from 100 to 700~MeV. The angular resolution is 
of the order of 0.02 radians. The end-cap BGO calorimeter with a 
thickness of 13.4~X$_0$ placed inside the solenoid 
has energy and angular resolution varying from 9\% to 4\% and from 
0.03 to 0.02 radians respectively for the photon energy in the range 
100 to 700~MeV.
The barrel and end-cap calorimeter systems cover a solid angle of
$0.92\times4\pi$ radians. 

The experiment was performed in the c.m. energy range 360--1380~MeV. 
This analysis is based on the data sample corresponding to
integrated luminosity of 10.7~pb$^{-1}$ collected in 1997--2000 
in the energy range above the $\ompi$ threshold, in 
c.m. energy steps of about 5~MeV.
The beam energy spread is about 400~keV at 1000~MeV.
The luminosity is measured using events of Bhabha scattering 
at large angles~\cite{prep}.

\section{Data analysis}

Since the decay mode $\omega\to\pig$ has been chosen for analysis,
at the initial stage  events are selected which have no tracks 
in the DC, five or six photons, the total energy deposition
$E_{tot} > 1.5\, E_{beam}$, the total momentum 
$P_{tot} < 0.4\, E_{beam}$ and at least three photons detected in the
CsI calorimeter. The minimum photon energy is 30~MeV for the CsI and 
40~MeV for the BGO calorimeter. We select 3045 events after these
requirements.

Then a kinematic fit requiring energy-momentum conservation is
performed with the additional reconstruction of two $\pi^0$.
The reconstruction procedure assumes five photons, i.e. if more than 
five photons are found in the event, a combination of five photons 
with the minimum $\chi^2$ is chosen.
We require good reconstruction quality ($\chi^2 < 6$) and
the ratio of the reconstructed to measured energy to be 
$0.7 < \omega_i\, / E_i < 1.9$ for each photon. 2598 events remain at 
this stage.

Since the background from the $\phi$ meson decays is large in
the  energy range
$1010 < \sqrt{s} <1028$~MeV, we exclude this energy
range from our analysis. At other energies there are no sources of 
multiphoton events with a significant cross section.
The background from QED processes 
($\ee\to 3\gamma,4\gamma$)~\cite{kura}
is efficiently suppressed by the cut on the minimum number of photons. 
Another possible source of background is the process
$\ee\to K_S^0 K_L^0$, $K_S^0\to\pi^0\pi^0$. However, 
except for the $\phi$ meson energy range its cross
section is small~\cite{cmd_kskl} and is efficiently rejected by the
total energy deposition and momentum cuts. 
The processes $\ee\to\eta\gamma$, $\eta\to\pi^0\pi^0\pi^0$ and
$\ee\to\ompi\pi^0$,  $\omega \to \pig$, 
can contribute to the $\pipig$ final state if the two soft photons 
are not 
reconstructed. For both of them the values of the cross section beyond 
the $\phi$ meson energy range are also small~\cite{cmd_etag,cmd_ompipi}.
From the Monte Carlo (MC) study we expect that all above listed processes 
give negligible background to the final state studied ($\lesssim 1\%$).

All events meeting the selection criteria 
are considered to be from the process $\ee\to\pipig$.
They are subdivided into two classes: those from the $\ompi$
intermediate state are selected with a cut 
$|M(\pig)-M_\omega|<80$~MeV$/c^2$ (2382 events) and 
those from the non-$\ompi$ are obtained by inverting the
cut above (216 events). The $\pipig$ final state can also arise from
the process $\ee \to \rho^0\pi^0$ followed by the 
$\rho^0 \to \pi^0 \gamma$ decay. This contribution estimated
from the measured cross section of the  process 
$\ee \to \rho\pi \to \pi^+\pi^-\pi^0$~\cite{snd_3pi} and 
$\mathcal{B}(\rho^0 \to \pi^0\gamma)$~\cite{pdg} 
appears to be about  $10^{-3}$ of the total one and can be neglected.

\begin{figure}
  \includegraphics[width=0.5\textwidth]{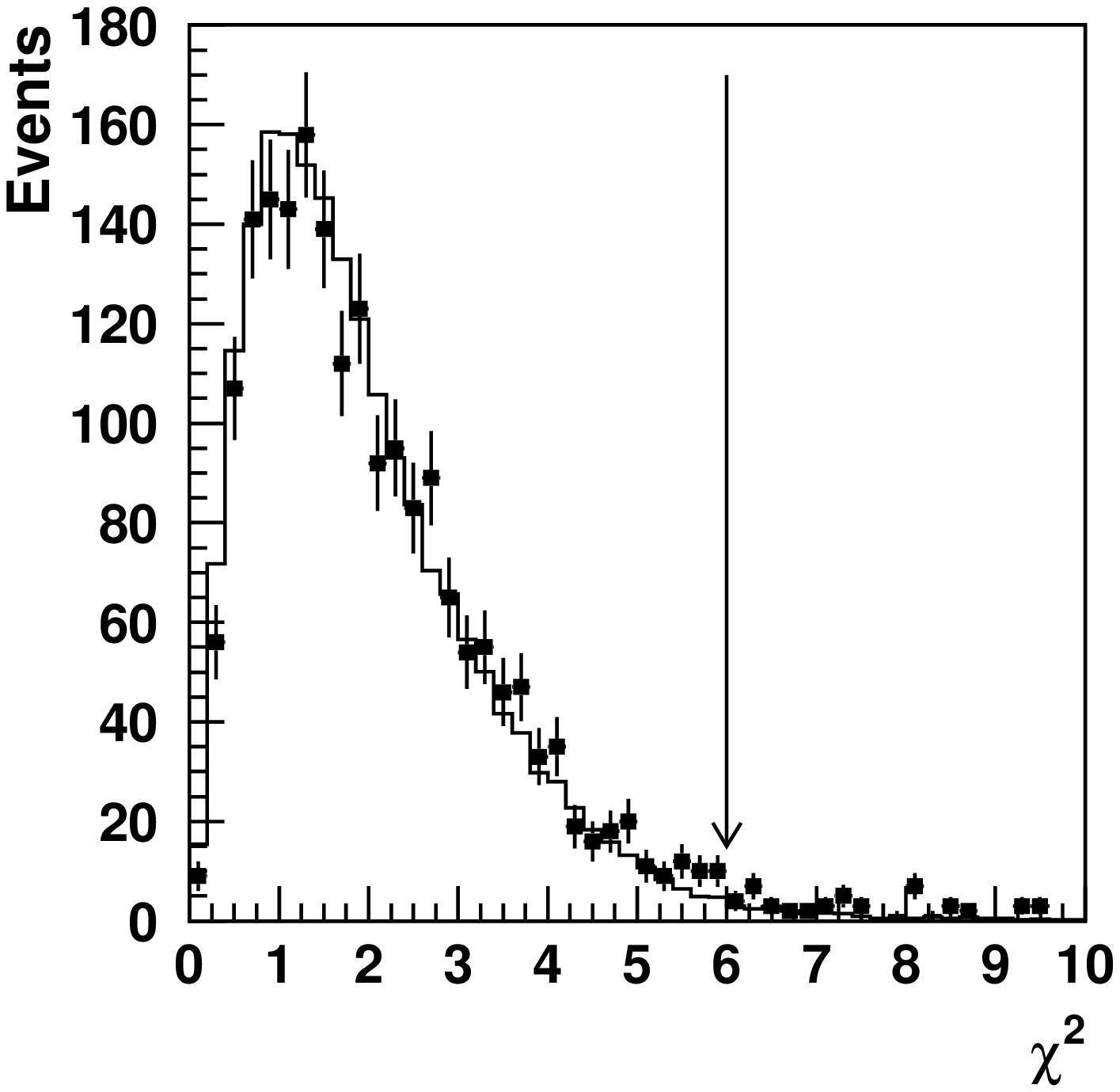}
  \includegraphics[width=0.5\textwidth]{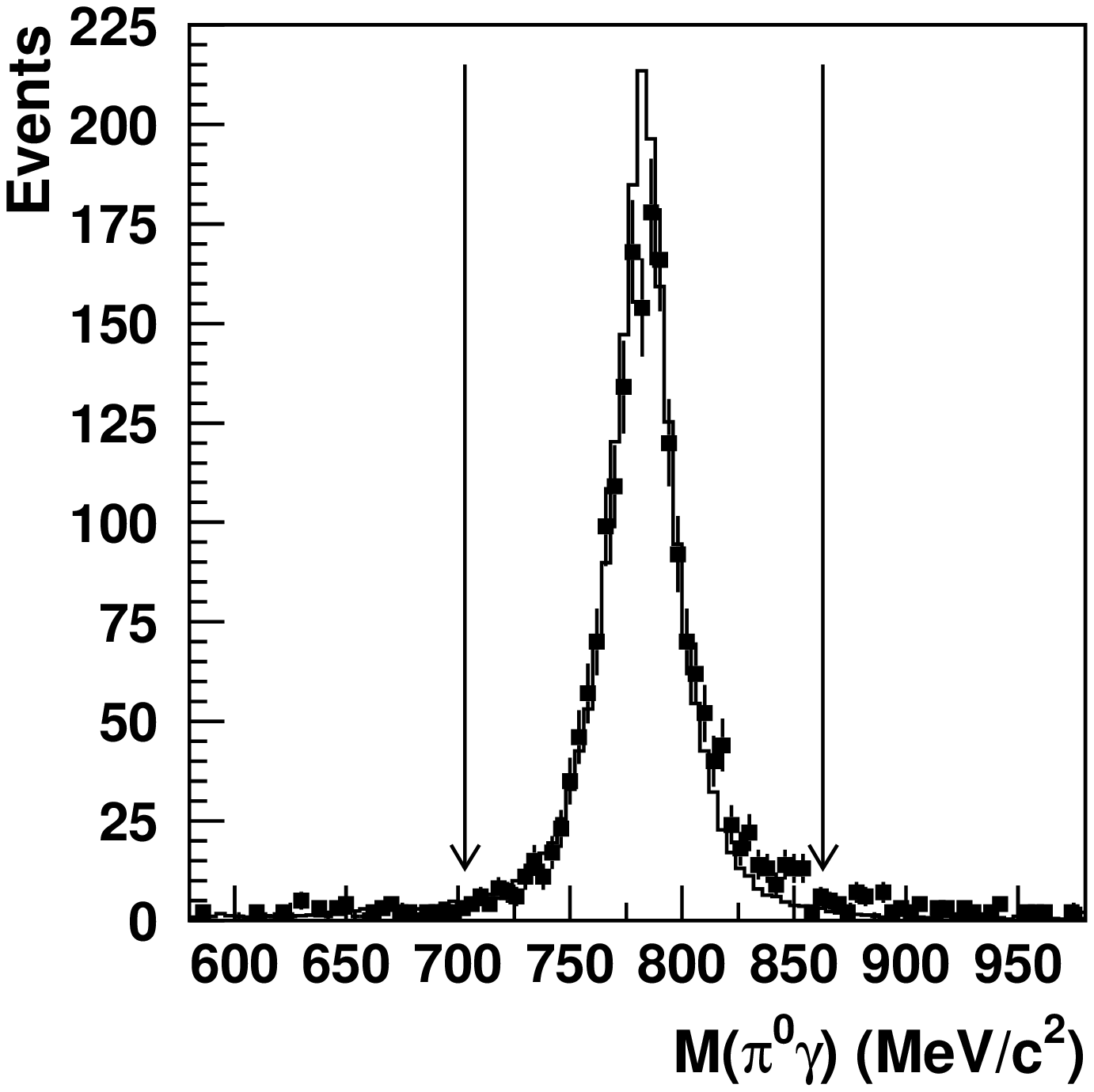}
  \caption{The $\chi^2$ distribution (left) and $\pig$
  invariant mass spectrum (right). The points with errors represent
    experimental events and the histograms show MC simulation. The
    arrows indicate the cuts imposed.}
  \label{om_comp}
\end{figure}

Figure~\ref{om_comp} shows the $\chi^2$ 
distribution for the events with 
$|M(\pig)-M_\omega|<80$~MeV$/c^2$ and the $\pig$
invariant mass distribution for the events with $\chi^2<6$.

\begin{figure}
\centering
  \includegraphics[width=0.5\textwidth]{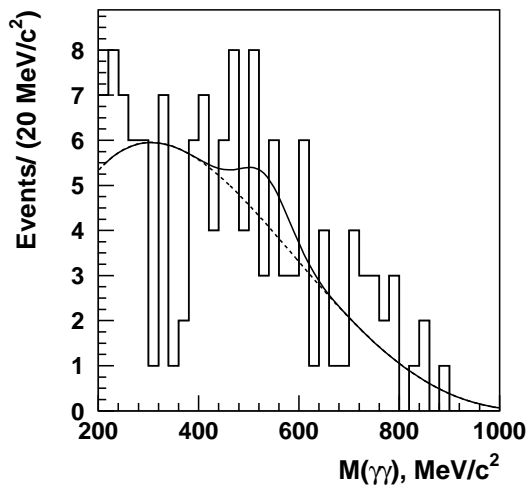}
  \caption{The $M_{\gamma\gamma}$ distribution for the $\etapig$
    candidates. The histogram represents experimental events and 
    the solid curve shows the fit result. The dashed curve corresponds 
    to the background contribution.}
  \label{mgg}
\end{figure}

For a search of events of the process $\ee\to\etapig$  we first apply
the same criteria as for the process $\ee\to\pipig$ at the initial stage. 
After that a kinematic fit requiring energy-momentum conservation is
performed with the additional reconstruction of one soft $\pi^0$.
We require good reconstruction quality, $\chi^2 < 6$.
To reject the dominant background from the process
$\ee\to\ompi\to\pipig$,
we perform an additional kinematic fit with the  $\pipig$ hypothesis and
reject events that are consistent with it,
$\chi^2_{\pipig}<6$. Then we look for a possible $\eta$ signal
in the invariant mass of any two photons of the remaining
three, $M_{\gamma\gamma}$.
The $M_{\gamma\gamma}$ distribution is approximated with a
Gaussian for the signal and polynomial function for the background.
The Gaussian mean value and width are fixed from the MC simulation of 
the signal events. The background shape is obtained using the  
$\ompi\to\pipig$ MC. In all energy ranges the resulting $\etapig$ signal 
is consistent with zero. Figure~\ref{mgg} shows the  
$M_{\gamma\gamma}$  distribution for all the selected events. 

\section{Results}
\subsection{Approximation of the cross sections}

At each energy point the cross section of the process $\sigma$ is 
calculated from the observed number of events $N_{\ompi}$ by using the 
following formula:
\begin{equation}
\sigma(\sqrt{s})=\frac{N_{\ompi}(\sqrt{s})}
{L(\sqrt{s})\cdot\varepsilon(\sqrt{s})\cdot(1+\delta(\sqrt{s}))}
  \label{Nth}
\end{equation}
where $L$ is the integrated luminosity at the c.m. energy 
$\sqrt{s}=2E_{beam}$, 
$\varepsilon$ is the detection efficiency and
$(1+\delta)$ is the radiative correction at the corresponding energy.  

To calculate the detection efficiency
we use Monte Carlo simulation taking into account the neutral 
trigger (NT) efficiency. NT is based on the information from the CsI 
calorimeter and its efficiency depends on the number of clusters and 
total energy deposition. The NT efficiency is estimated using events of
the processes $\ee\to\pi^+\pi^-\pi^0$ and $\ee\to\pi^+\pi^-\pi^0\pi^0$. 
We require  
the charged trigger signal and three or more clusters in the CsI 
calorimeter, and study the NT efficiency as a function of the energy 
deposition in CsI. The NT efficiency varies from 92\% at 920~MeV to 
about 98\% at 1380~MeV~\cite{mast}.

\begin{table*}
\caption{The energy, integrated luminosity, detection efficiency, 
  number of selected events, radiative correction, Born cross
  section, vacuum polarization correction and ``bare'' cross section 
of the process $\ee\to\ompi\to\pipig$.}
\medskip
\begin{tabular*}{\textwidth}{c@{\extracolsep{\fill}}ccccccc}
\hline\hline
$\sqrt{s}$, MeV & \hspace*{2mm}$L$, nb$^{-1}$& $\varepsilon$, \% & $N_{\ompi}$& 
$1+\delta$& $\sigma$, nb & $|1 - \Pi(s)|^2$ & $\hat{\sigma}$, nb\\\hline
  920&  458& 13.2&   3& 0.857& $0.06\pm 0.03$& $0.964$& $0.06\pm 0.03$\\
  940&  328& 18.0&   6& 0.833& $0.12\pm 0.04$& $0.966$& $0.12\pm 0.04$\\
  950&  226& 18.7&  13& 0.840& $0.37\pm 0.09$& $0.968$& $0.35\pm 0.08$\\
  960&  250& 18.9&  17& 0.849& $0.42\pm 0.09$& $0.970$& $0.41\pm 0.08$\\
  970&  250& 19.2&  22& 0.857& $0.54\pm 0.10$& $0.972$& $0.52\pm 0.10$\\
  984&  430& 19.5&  43& 0.867& $0.59\pm 0.08$& $0.977$& $0.58\pm 0.08$\\
 1004&  476& 20.0&  53& 0.879& $0.63\pm 0.08$& $0.995$& $0.63\pm 0.08$\\
 1034&  406& 20.6&  82& 0.893& $1.10\pm 0.11$& $0.931$& $1.02\pm 0.10$\\
 1044& 1000& 20.8& 172& 0.897& $0.92\pm 0.06$& $0.944$& $0.87\pm 0.06$\\
 1061&  543& 21.2&  86& 0.902& $0.83\pm 0.08$& $0.953$& $0.79\pm 0.08$\\
 1083&  320& 21.6&  77& 0.906& $1.23\pm 0.13$& $0.958$& $1.18\pm 0.12$\\
 1103&  290& 22.0&  50& 0.910& $0.86\pm 0.11$& $0.960$& $0.83\pm 0.11$\\
 1123&  308& 22.3&  80& 0.913& $1.27\pm 0.13$& $0.962$& $1.23\pm 0.13$\\
 1142&  302& 22.6&  74& 0.916& $1.18\pm 0.13$& $0.963$& $1.14\pm 0.12$\\
 1163&  293& 23.0&  88& 0.917& $1.43\pm 0.14$& $0.964$& $1.37\pm 0.13$\\
 1183&  485& 23.3& 131& 0.918& $1.26\pm 0.10$& $0.965$& $1.22\pm 0.10$\\
 1204&  301& 23.6&  85& 0.919& $1.30\pm 0.13$& $0.965$& $1.26\pm 0.13$\\
 1226&  249& 23.9&  71& 0.920& $1.30\pm 0.14$& $0.966$& $1.25\pm 0.14$\\
 1246&  348& 24.1& 121& 0.920& $1.57\pm 0.13$& $0.966$& $1.52\pm 0.13$\\
 1266&  417& 24.4& 138& 0.920& $1.47\pm 0.12$& $0.967$& $1.43\pm 0.11$\\
 1286&  497& 24.6& 152& 0.921& $1.35\pm 0.10$& $0.967$& $1.31\pm 0.10$\\
 1304&  499& 24.7& 159& 0.921& $1.40\pm 0.10$& $0.967$& $1.35\pm 0.10$\\
 1326&  541& 24.9& 176& 0.924& $1.41\pm 0.10$& $0.968$& $1.37\pm 0.10$\\
 1345&  428& 25.1& 140& 0.926& $1.41\pm 0.11$& $0.968$& $1.36\pm 0.11$\\
 1364&  499& 25.2& 177& 0.930& $1.51\pm 0.11$& $0.968$& $1.46\pm 0.10$\\
 1380&  536& 25.3& 166& 0.935& $1.31\pm 0.10$& $0.969$& $1.27\pm 0.09$\\
\hline\hline
\end{tabular*}
\label{ompi_cs}
\end{table*}

\begin{figure*}
\centering
  \includegraphics[width=0.49\textwidth]{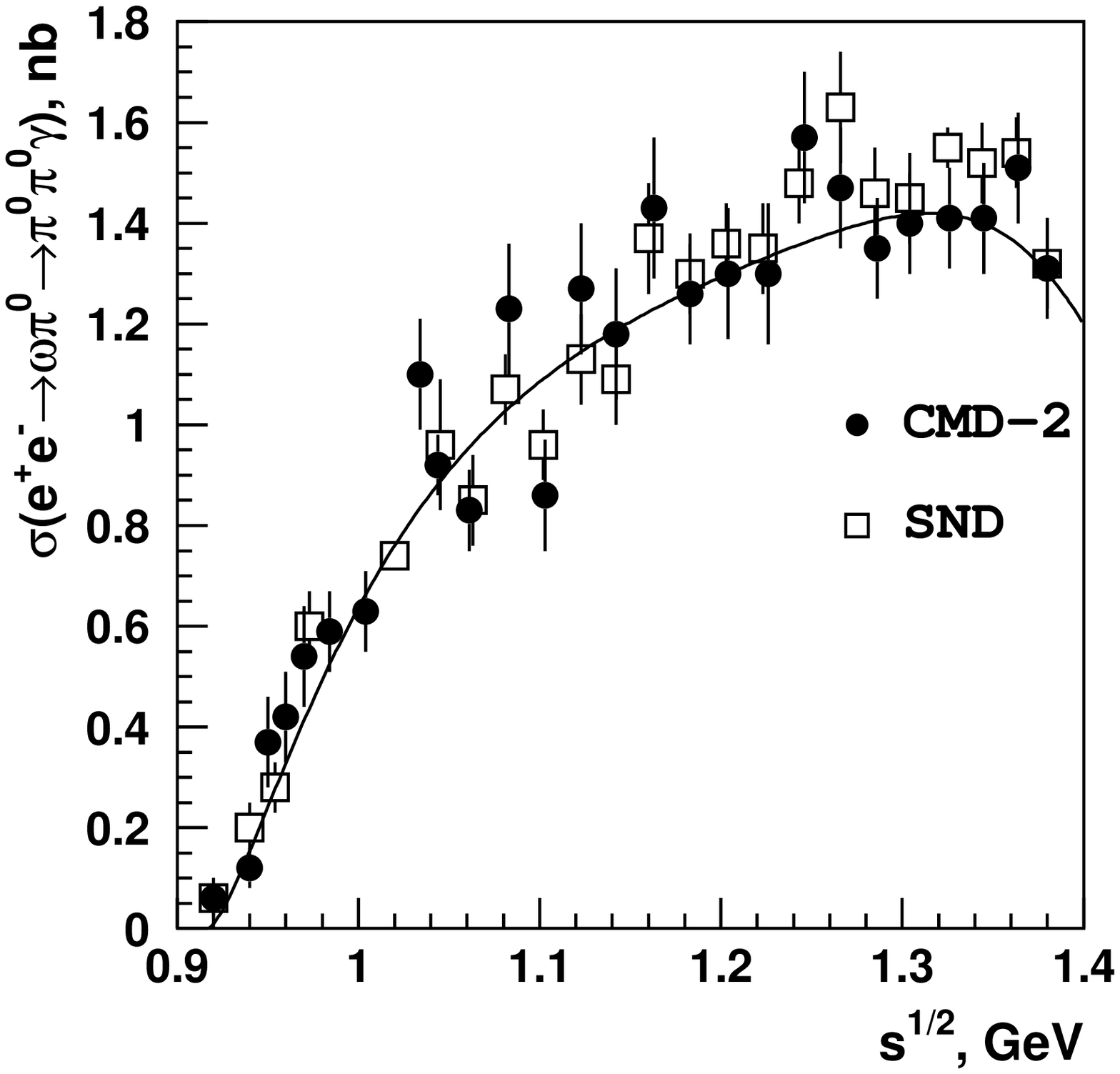}\hfill
  \includegraphics[width=0.49\textwidth]{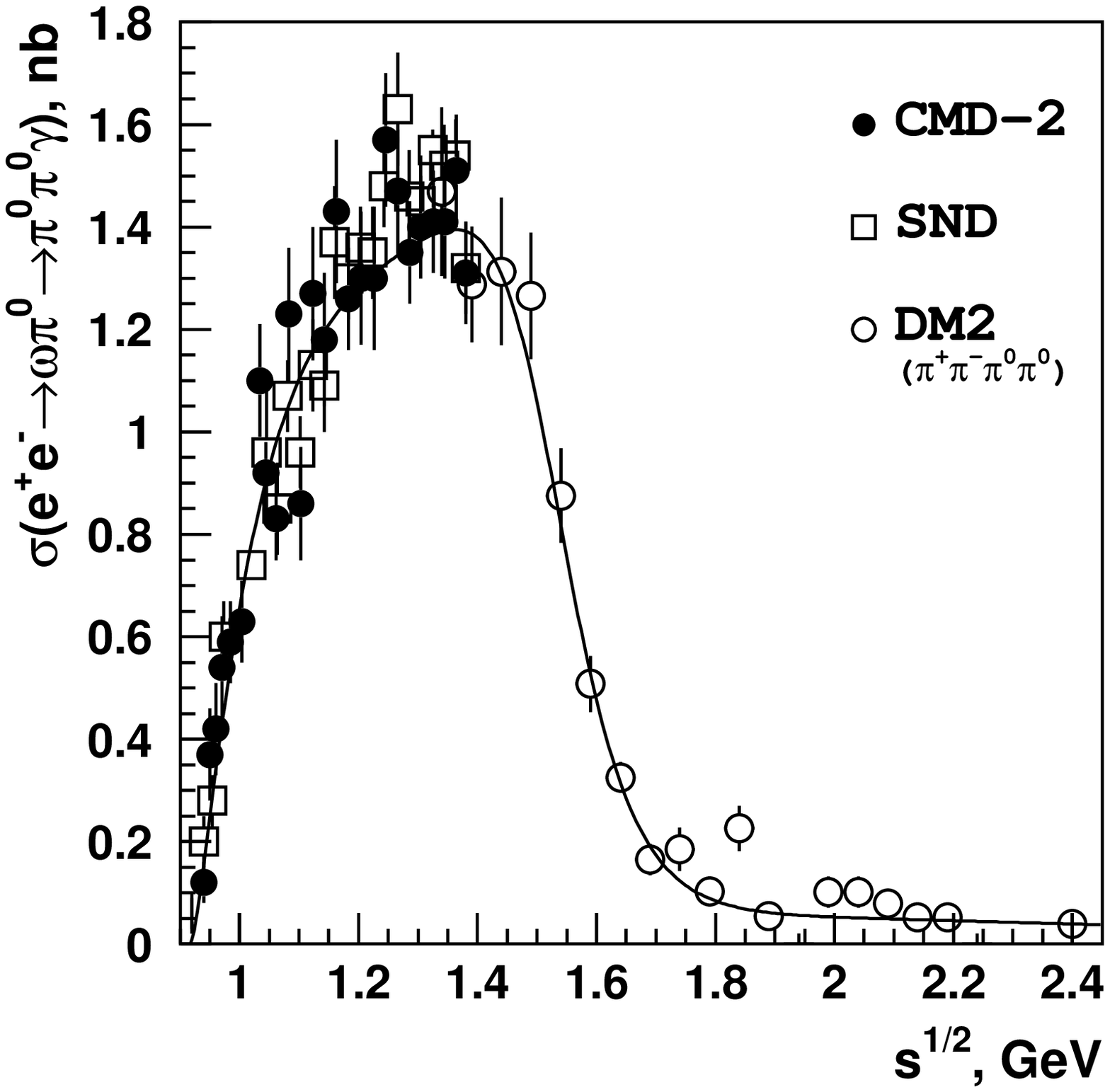}
  \caption{The cross section of the process
  $\ee\to\ompi\to\pipig$. The results of CMD-2 (this work), 
  SND~\cite{snd_ompi} and DM2~\cite{dm2_ompi} are shown. 
  The curves are the results: a fit to the CMD-2 data only 
  (Fit I, left) and a combined fit to the CMD-2 and DM2 data 
(Fit II, right).}
  \label{cs_cmd}
\end{figure*}

The obtained Born cross section of the process 
$\ee \to \ompi \to \pipig$ is shown in  Fig.~\ref{cs_cmd}
while Table~\ref{ompi_cs} lists detailed information on the analysis
of this reaction.
It is this cross section (the ``dressed'' one from the column VI) 
that should be used in
the approximation of the energy dependence with resonances. For
applications to various dispersion integrals like that for the leading
order hadronic contribution to the muon anomalous magnetic moment, one
should use the ``bare'' cross section. 
Following the procedure in
Ref.~\cite{rho}, the latter is obtained from the ``dressed'' one by
multiplying it by the vacuum polarization correction $|1 - \Pi(s)|^2$,
where  $\Pi(s)$ is the photon polarization operator calculated taking
into account the effects of both leptonic and hadronic vacuum
polarization. The value of the correction and the ``bare'' cross
section $\hat{\sigma}$ are presented in two last columns of 
Table~\ref{ompi_cs}. 

The maximum likelihood method is applied to fit the experimental
data to the relation (\ref{Nth}) with the parameterization of the
cross section described below. 
The radiative corrections are calculated during the fit according 
to~\cite{radcor}.
The dependence of the detection efficiency on the energy 
of the emitted photon is determined from simulation.

The Born cross section of the process can be written as:
\begin{equation}
  \sigma_0(s)=\frac{4\pi\alpha^2}{s^{3/2}}
  \Bigl(\frac{g_{\rho\omega\pi}}{f_\rho}\Bigr)^2
  \Bigl|\frac{m^2_\rho}{D_\rho}+
  A_1 \frac{m^2_{\rhop}}{D_{\rhop}}+
  A_2 \frac{m^2_{\rhopp}}{D_{\rhopp}}\Bigr|^2 \cdot P_f(s)\; .
\label{ompi_rho}
\end{equation}
Here $g_{\rho\omega\pi}$ is the coupling constant of the transition
$\rho\to\omega\pi$;
the coupling constant $f_\rho$ is calculated from the $\rho\to\ee$ decay
width: $\Gamma_{\rho ee}=4\pi m_\rho \alpha^2/(3f^2_\rho)$;
$m_V$ is the mass and $D_V$ is the propagator of the vector meson 
$V$ given by $D_V(s)=s-m^2_V+i\sqrt{s}\Gamma_V(s)$, $\Gamma_V$ is the
corresponding width.
The real parameter $A_1=g_{\rhop\omega\pi}/g_{\rho\omega\pi}\cdot
f_\rho/f_{\rhop}$ is the ratio of the coupling constants for
the $\rho$ and $\rhop$ mesons while $A_2$ is similarly defined
for the $\rhopp$ meson. The factor $P_f(s)$ describes the energy
dependence of the decay width into the $\ompi$ final state.
For the infinitely narrow $\omega$ resonance
$P_f(s)=1/3\cdot p^3_\omega\cdot {\mathcal B}_{\omega\to\pig}$,
where $p_\omega$ is the $\omega$ meson momentum. However, this expression
is not valid in the narrow region near the $\ompi$ threshold where 
we take into account the finite width of the $\omega$ meson.
 
\subsection{Results of the fits} 

In all the following fits $g_{\rho\omega\pi}$ and $A_1$ are free 
parameters. 
We perform three main fits: to CMD-2 data only (Fit I) and two 
combined fits to our data and those of the DM2 
measurement~\cite{dm2_ompi}. A free correction factor for the DM2 data 
has been included in the combined fits to take into account  
systematic uncertainties of both DM2 and our measurements.
 
For the $\rho$ resonance the energy dependence of the total width
is described by the $\pipi$ and $\ompi$ decay modes:
$$
\Gamma_{\rho}(s)=\Gamma_{\rho}(m_\rho^2) \frac {m^2_{\rho}}{s}    
\Bigl(\frac {p_{\pi}(s)} {p_{\pi}(m^2_{\rho})}\Bigr)^3 +
\frac{g^2_{\rho\omega\pi}}{12\pi} p^3_{\omega}(s)\ .
$$

In the first fit we fix the $\rhop$ meson mass and width at their world 
average values~\cite{pdg} since our data cover the energy range below 
1380~MeV or the left slope of the $\rhop$ only.
We also neglect the contribution of the $\rhopp$  so that
the corresponding coupling constant $A_2=0$.
In the second fit the DM2 data above 1400~MeV allow a 
determination of the $\rhop$ parameters directly from the fit. For the
$\rhop$ resonance $\Gamma(s)$ is also described by two decay 
channels: $\pipi$ and $\ompi$.
We checked that possible contributions of the
$K\bar{K}$, $\eta\pi\pi$ and $a_1(1260)\pi$ channels to the 
$\rhop$ width  only slightly affect the results of the fits. 
That is clear since the contribution of the two former is 
numerically small whereas the latter has energy dependence 
proportional to the first power of momentum and produces no effect 
compared to the fast growing term corresponding to the 
$\ompi$ state. 
In contrast to the previous case, the relative 
probabilities of the $\pipi$ and $\ompi$ modes
are unknown so that the expression for the width is written as
$$\Gamma_{\rhop}(s)=
\Gamma_{\rhop}(m^2_{\rhop}) \Bigl[ {\mathcal B}_{\rhop \to \ompi}  
\Bigl(\frac {p_{\omega}(s)} {p_{\omega}(m^2_{\rhop})}\Bigr)^3 +
(1-{\mathcal B}_{\rhop \to \ompi})\frac {m^2_{\rhop}} {s}    
\Bigl(\frac {p_{\pi}(s)} {p_{\pi}(m^2_{\rhop})}\Bigr)^3\Bigr] \,
$$ 
where ${\mathcal B}_{\rhop \to \ompi}$ is the branching ratio of the
$\rhop \to \ompi$ decay. Its value was varied from zero to unity
to estimate the model uncertainty.  As in the 
Fit I, $A_2=0$. In the third fit both $\rhop$ and $\rhopp$ can decay into 
$\ompi$ with  the $\rhop$ parameters free and the $\rhopp$  mass 
and width fixed~\cite{pdg}.

Results of the fits are shown in Table~\ref{BB} and in 
Fig.~\ref{cs_cmd} by the curves.
All the fits describe data of CMD-2 well. A large increase of 
$\chi^2/$n.d.f. for the Fits II and III comes from two DM2 points at 
1.84 and 1.94~GeV only. Without these points $\chi^2/$n.d.f. = 43/38 
for the second fit and 42/37 for the third one.  
The values of the correction factor for the DM2 data obtained 
from the fits agree with their estimated systematic uncertainty of 15\%. 

\begin{table*}
\caption{ The fit results in various models}
\medskip
\begin{tabular*}{\textwidth}{l@{\extracolsep{\fill}}ccc}\hline\hline
Fit parameters      & Fit I  & Fit II & Fit III \\
\hline
$g_{\rho\omega\pi}$, GeV$^{-1}$ & $16.7\pm 0.4\pm 0.6$ & 
$17.0\pm 0.4\pm 0.6$ & $16.7\pm 0.4\pm 0.6$\\
\hline
$A_1$, $10^{-2}$ & $-8.0\pm 0.8\pm 1.0$ & 
$-8.5\pm 0.9\pm 1.0$ & $-10.0\pm 1.6\pm 1.0$\\
\hline
$M_{\rhop}$, GeV       & $\equiv 1450$& $1564\pm 9\pm 25$ &
$1582\pm 17\pm 25$\\
\hline
$\Gamma_{\rhop}$, GeV  & $\equiv 310$ & $390\pm 36\pm 10$ &
$429\pm 42\pm 10$\\
\hline
$A_2$, $10^{-2}$   & $\equiv 0$ & $\equiv 0$ & $0.7\pm 0.6$\\
\hline
$M_{\rhopp}$, GeV      & --- & --- & $\equiv 1700$\\
\hline
$\Gamma_{\rhopp}$, GeV & --- & --- & $\equiv 240$\\
\hline
Correction factor & --- & $1.18\pm 0.07$& $1.20\pm 0.07$\\\hline
$\chi^2/$n.d.f.         & 23 / 23       & 62 / 40 & 60 / 39\\
\hline\hline
\end{tabular*}
\label{BB}
\end{table*}

The value of $g_{\rho\omega\pi}$ is consistent within errors with 
the experimental values from 12 to 17~GeV$^{-1}$ following from the 
$\omega \to \pig$, $\rho \to \pig$ and 
$\omega\to \rho\pi \to  \pi^+\pi^-\pi^0$ decays.  It is also 
consistent with the theoretical estimates based on the QCD sum 
rules predicting 
the broad range from 9 to 16~GeV$^{-2}$~\cite{qcdsr}. 
The measured cross section agrees with the previous measurement 
of the SND group~\cite{snd_ompi}. 

It is interesting to note that the value of $A_1$, the relative 
weight of the $\rhop$ amplitude, within the errors is consistent
with the corresponding value obtained in the CMD-2 analysis of the
reaction $\ee\to\pipi$ in the vicinity of the $\rho$ 
meson~\cite{rho}. This is a natural consequence of the fact
that the isovector component of the electromagnetic current is
the same in all reactions independently of the specific final state.  
 
From Fig.~\ref{cs_cmd} it is clear that the DM2 data above 1600~MeV
do not require another ($\rho(1700)$) resonance. As follows from the 
Fit III, an attempt to add such a state to the cross 
section parameterization does not improve the fit and its contribution
is compatible with zero. This is consistent with the conclusion of 
the DM2 group~\cite{dm2_ompi}.

\subsection{Search for direct processes}

To estimate a possible contribution to the cross section of the
process $\pipig$ from the non $\ompi$ intermediate state, we used the
following procedure.
We compare the observed number of the  non $\ompi$ events 
($|M(\pig)-M_\omega|>80$~MeV$/c^2$) 
with the MC expectation for the $\ompi$  final state. The results of this
analysis are shown in Table~\ref{non-ompi}. The observed number of
events somewhat exceeds the expectation, but doesn't contradict 
it within errors. All distributions for these events are compatible
with those expected for the $\ompi$, so the observed excess of events can
be due to the imperfect Monte Carlo simulation. From the difference
between the observed and expected number of events we set the 90\% CL 
upper limits for the non $\ompi$ cross section in various energy ranges. 
To calculate the detection efficiency, an $f_0(600)\gamma$ 
intermediate mechanism interfering with the $\ompi$ was assumed.
The contribution of the non $\ompi$ process to the $\omega$
selection criteria averaged over the whole energy range is estimated 
to be less than 3\% of the $\ee\to\ompi\to\pipig$ cross section.

\begin{table*}
\caption{The energy, integrated luminosity, efficiency, number of 
  observed events, expectation from $\ompi$ MC and 90\% CL UL for the 
  cross section of the non $\ompi$ cross section}
\medskip
\begin{tabular*}{\textwidth}{c@{\extracolsep{\fill}}ccccc}
\hline\hline
$\sqrt{s}$, MeV & $L$, nb$^{-1}$ & $\varepsilon$, \% & $N_{\pipig}$ & 
$N_{MC}$ & $\sigma$, nb (90\% CL) \\
\hline
 920--1004 & 2418 &  9.0 & 33 & 28.1 & 0.07\\
1034--1200 & 3947 &  7.9 & 88 & 70.4 & 0.11\\
1200--1300 & 1812 &  9.5 & 40 & 34.8 & 0.09\\
1300--1380 & 2503 & 10.5 & 55 & 49.7 & 0.07\\
\hline\hline
\end{tabular*}
\label{non-ompi}
\end{table*}

As noted earlier, we do not observe an $\etapig$ signal and assuming an
$a_0(980)\gamma$ intermediate mechanism  can set upper limits
for the cross section of the process $\ee \to \etapig$. 
Table~\ref{etapi0g} shows results of this study. The listed 
efficiencies include the $\eta\to\gamma\gamma$ branching 
fraction~\cite{pdg}.

\begin{table*}
\caption{The energy, integrated luminosity, efficiency, 90\% CL UL for 
  $\etapig$ events and 90\% CL UL for the cross section of the 
  $\etapig$ cross section}
\medskip
\begin{tabular*}{\textwidth}{c@{\extracolsep{\fill}}cccc}
\hline\hline
$\sqrt{s}$, MeV & $L$, nb$^{-1}$ & $\varepsilon$, \% & $N_{\etapig}$ & 
$\sigma$, nb  (90\% CL) \\ 
\hline
 920--1004 & 2418 & 2.2 & 6.9 & 0.13\\
1034--1200 & 3947 & 3.2 & 7.7 & 0.06\\
1200--1300 & 1812 & 3.3 & 8.7 & 0.14\\
1300--1380 & 2503 & 3.1 & 7.6 & 0.10\\\hline\hline
\end{tabular*}
\label{etapi0g}
\end{table*}
   
\subsection{Systematic errors}

The main sources of  systematic uncertainties in the cross section 
determination are listed in Table~\ref{syst}.
The systematic error due to selection criteria is obtained by varying
the photon energy threshold, total energy deposition, total momentum,
$\chi^2$ and $M(\pig)$ cuts.
The uncertainty in the determination of the integrated luminosity 
comes from the selection criteria of Bhabha events, radiative
corrections and calibrations of DC and CsI. The error of the NT 
efficiency was estimated by trying various
fitting functions of the energy dependence and variations of the
cluster threshold. The uncertainty of the radiative corrections comes 
from the dependence on the emitted photon energy and the accuracy of
the theoretical formulae. The resulting systematic uncertainty of the 
cross section quoted in Table~\ref{syst} is 6.6\%.

\begin{table*}
\caption{Main sources of systematic errors}
\medskip
\begin{tabular*}{\textwidth}{l@{\extracolsep{\fill}}c}
\hline\hline
Source & Contribution, \%\\
\hline
Selection criteria & 5\\
Non $\ompi$ contribution& 3\\
Luminosity & 2\\
Trigger efficiency & 2 \\
Radiative corrections & 1 \\
\hline
Total & 6.6\\
\hline\hline
\end{tabular*}
\label{syst}
\end{table*}

\section{Discussion}

As can be seen from Fig.~\ref{cs_cmd}, the cross section of the
process $\ee\to\ompi \to \pipig$ grows fast with energy in 
the whole energy range covered by
the CMD-2 and reaches its maximum at 1.35--1.40~GeV. If we divide
the value of the cross section by the branching ratio 
$\omega\to\pig$, the resulting cross section is consistent
with that from the $\omega\to\trpi$ channel within the
experimental uncertainties~\cite{cmd2_ompi,snd1}.
At higher energy,  the cross section starts 
falling rapidly as the DM2 results show. The whole pattern
of the energy dependence is well described by the interference of the
$\rho$ and $\rhop$ resonances and, as noted above, hardly requires
a third resonance, $\rhopp$. 

We do not perform here a detailed CVC test by comparing our results 
on the cross section with the values of the spectral functions of 
the $\omega\pi$ final state in $\tau$
lepton decays measured by ALEPH~\cite{tau1} and CLEO~\cite{tau2}. As
noted recently~\cite{cvc}, significant disagreement between the
spectral functions of the $4\pi$ final state  observed now 
requires a serious analysis of both data sets as well as of the
necessary SU(2) breaking corrections. However, from 
various mass distributions in $\ee\to 4\pi$ and 
$\tau^- \to 4\pi\nu_{\tau}$ it is clear that 
qualitatively the mechanisms of the $4\pi$ decay of  
the $\tau$ lepton and corresponding process in $\ee$
annihilation are very close~\cite{bond}. Therefore, it is interesting 
to note that high statistics analysis of the $\omega\pi$ component
in $\tau$ decays performed by CLEO confirms our conclusion that
only one $\rho$-like resonance is needed for an acceptable description of
the data. Its mass ranging between 1520 and 1630~MeV and large
width of 400--650~MeV strongly depend on the parameterization of 
the energy dependence of the width and are consistent with our results
in Table~\ref{BB}. Recently CLEO observed a strong $\omega\pi$
component in the $4\pi$ state produced together with $D^{(*)}$ in $B$
decays~\cite{cleob}. Its spin parity analysis shows a preference    
for a wide $1^{--}$ resonance with a mass of $1349\pm 27$~MeV and 
width of $547\pm 98$~MeV identified as the $\rho(1450)$. 
Their data do not require a higher, $\rho(1700)$ resonance decaying
into the $\ompi$ final state. 

It is well known that the energy range under study can not be
described now from the first principles and one has to use the
predictions of specific models. In the popular relativized quark
model with chromodynamics it is predicted that if 
the $\rho(1450)$ and $\rho(1700)$ are
the $2^3S_1$ and $1^3D_1$ $q\bar{q}$ states respectively, only the 
former has large probability of decay into
$\ompi$~\cite{gi}. This is consistent with our observation as
well as with the data of CLEO on $\tau$ and $B$ decays described above.
However, the mentioned above model also predicts strong suppression of the
$a_1(1260)\pi$ mode and dominance of the $h_1\pi$ mode for 
the $4\pi$ decays of the $2^3S_1$ quarkonium. This is in strong
contrast with the observations of CMD-2~\cite{cmd2_ompi} and 
CLEO~\cite{tau2} that support the $a_1(1260)\pi$ 
dominance\footnote{An $a_1(1260)\pi$ component of the 4$\pi$ final
state, although not as significant, has also been observed by Crystal 
Barrel in $\bar{\rm p}{\rm n}$ annihilation at rest~\cite{cbal}.}. 
Therefore, it was argued in Ref.~\cite{hybrid1} that to reconcile
the data, a hybrid component is needed. The model calculations
of various decay modes of the hybrids show that the $a_1(1260)\pi$ is
the dominant mode, and the  $\omega\pi$ is also 
significant~\cite{hybrid3}. Thus, one can not exclude that the observed
picture is in fact more complicated and one or several vector hybrids
exist in the energy range between 1 and 2~GeV in addition to
the $\rho', \omega', \phi'$. If masses of the hybrid states
are close to those of quarkonia and they have common decay channels,
their interference could produce a peculiar energy dependence of the 
cross sections.    

Our results can be also used to set upper limits for various
radiative decays of the higher vector mesons to scalar and tensor
mesons predicted in some models~\cite{kalash}.
In Table~\ref{tab:rad} we present the upper 
limits for the corresponding cross section in the highest energy range
accessible to our experiment --- from 1300 to 1380 MeV. The choice of
this energy range is clear if one takes into account that the masses
of the majority of the considered scalar and tensor
mesons are rather high so that the available phase space is limited. To
obtain the quoted limits we used the results from
Tables~\ref{non-ompi} and \ref{etapi0g} corrected for the 
branching ratio to the $\pi^0\pi^0~(\eta\pi^0)$ and the width of
the meson. Interference with the $\ompi$ intermediate mechanism was taken 
into account while calculating the detection efficiency.  

\begin{table*}
\caption{Upper limits for the cross section of the radiative
transition to the meson X}
\medskip
\begin{tabular*}{\textwidth}{c@{\extracolsep{\fill}}cc}
\hline\hline
Final state & Meson X & $\sigma$, nb  (90\% CL) \\ 
\hline
$\etapig$ & $a_0(980)$  & 0.05 \\
          & $a_2(1320)$ & 0.70 \\
\hline
$\pipig$ & $f_0(600)$  & 0.19 \\
         & $f_0(980)$  & 0.05 \\
         & $f_2(1270)$ & 0.23 \\
\hline\hline
\end{tabular*}
\label{tab:rad}
\end{table*}

Finally, we can use our results on the search for the $\etapig$ events 
to look for the $\omega(1650) \to \omega\eta$ decay that was
observed before in $\pi^- p$ collisions~\cite{eug}. In our case
the decay $\omega \to \pig$ leads to the $\etapig$ final
state. The fit of the observed events between 1300 and 1380 MeV
to the expected energy dependence gives:
$${\mathcal B}(\omega(1650)\to\ee)\cdot{\mathcal B}
  (\omega(1650)\to\omega\eta)< 6\times 10^{-6}$$  
at 90\% confidence level.

One can expect further significant improvements in our understanding of 
the light quark resonance spectroscopy in the vector meson sector 
when experiments at the upgraded collider VEPP-2000 begin
in Novosibirsk~\cite{vep2000}.

From the obtained upper limits for the cross section of the radiative
processes $\ee\to X\gamma$, $X\to\pi^0\pi^0, \eta\pi^0$ one can estimate
a possible contribution of the previously unstudied radiative
processes to the leading order hadronic correction to the muon
anomalous magnetic moment. Taking into account that a possible
contribution from the process $\ee \to \pi^+\pi^-\gamma$ is twice
that of $\ee\to \pipig$, one obtains  
$$
 {\rm a}^{LO,rad}_{\mu} < 0.45 \cdot 10^{-10}~~{\rm at}~~90\%\ {\rm CL}.
$$ 
This limit is negligible (less than 7\%) compared to the current 
uncertainty of ${\rm a}^{LO,had}_{\mu}$~\cite{cvc}.

\section{Conclusions}

The following results are obtained in this work:
\begin{itemize}
\item
Using a data sample corresponding to integrated luminosity of 
10.7~pb$^{-1}$, the cross section of the process $\ee\to\ompi\to\pipig$ 
has been measured in the c.m. energy range 920--1380~MeV. The
values of the cross section are consistent with those obtained by
the SND detector~\cite{snd_ompi}.
\item
The combined fit of the CMD-2 data and those from 
DM2 at higher energies confirms the existence of the 
$\rho(1450) \to \ompi$ decay mode while a significant 
$\rho(1700) \to \ompi$ decay is not needed to describe the data. 
\item
The 90\% CL upper limits for the cross sections of the direct  
$\ee\to\pipig, \etapig$ processes have been set for the studied 
energy range. It is shown that a possible contribution 
of such processes to ${\rm a}^{LO,had}_{\mu}$ is negligible.
\end{itemize}

The authors are grateful to the staff of VEPP-2M for the
excellent performance of the collider, and to all engineers and 
technicians who participated in the design, commissioning and operation
of CMD-2. We acknowledge useful and stimulating discussions with 
Yu.S.~Kalashnikova and E.S.~Swanson. This work is supported in part 
by the US Department of Energy, US National Science Foundation and
the Russian Foundation for Basic Research.

\end{document}